\documentclass[12pt]{article}

\usepackage[utf8]{inputenc}
\usepackage[english]{babel}
\usepackage{textcomp}
\usepackage{amsfonts}
\usepackage{amsmath}
\usepackage{amssymb}
\usepackage{mathrsfs}
\usepackage{graphics}
\usepackage{color}
\usepackage{upgreek}
\usepackage{epsfig}

\usepackage[numbers, sort&compress]{natbib}

\usepackage{slashed}

\usepackage{caption}
\usepackage{subcaption}

\usepackage[dvipsnames]{xcolor}

\begin{document}
\begin{titlepage}

\begin{center}
\mathversion{bold}
{\LARGE\bf
Non-minimal approximation for the type-I seesaw mechanism
\mathversion{normal}
}
\vskip .3cm
\end{center}
\vskip 0.5  cm
\begin{center}
{\large M.~N.~Dubinin}$^{1,2}$, 
{\large E.~Yu.~Fedotova}$^{1}$
\\
\vskip .7cm
{\footnotesize

$^{1}$ Skobeltsyn Institute of Nuclear Physics (SINP MSU), 
M.V. Lomonosov Moscow State University, Leninskie gory, GSP-1, 119991 Moscow, Russia\\[0.3cm]
$^2$ MegaScience Center, National University of Science and Technology MISIS,
Leninskiy Prospekt, 4, 119049, Moscow, Russia
\vskip .5cm
\begin{minipage}[l]{.9\textwidth}
\begin{center}
\textit{E-mail:}
\tt{dubinin@theory.sinp.msu.ru}, \tt{fedotova@theory.sinp.msu.ru}
\end{center}
\end{minipage}
}
\end{center}
\vskip 1cm
\begin{abstract}

A non-minimal approximation for effective masses of light and heavy neutrinos in the framework of a
type-I seesaw mechanism with three generations of sterile Majorana neutrinos 
is considered. 
The main results are: 
$(a)$ the next-order corrections to the effective mass matrix of heavy neutrinos due to terms ${\cal O}(\theta M_D)$ are obtained, which modify the  commonly used representation for the effective mass ($M_D$ is a Dirac neutrino mass when the electroweak symmetry is spontaneously broken); and
$(b)$ the general form of the mixing matrix is found in non-minimal approximation parametrized by a complex $3 \times 3$ matrix satisfying a nontrivial constraint. 
Numerical analysis within the $\nu$MSM framework demonstrates the very small effect of new contributions of 
direct collider observables as opposed to their possible significance for \mbox{cosmological models.} 

\end{abstract}
\end{titlepage}


\section{Introduction}
\label{intro}


Numerous consequences of the exceptionally successful standard model (SM) of elementary particle physics based on the gauge symmetry $SU(3)\times SU(2)_L\times U(1)$ have been confirmed by experiments on colliders and extracted particle beams. No significant deviations from the SM predictions were found, as a result of which it is considered as a mathematically consistent effective field theory for a very large energy scale. At the same time, it is obvious that SM is not a complete theory, since it cannot explain a number of observed phenomena not only in particle physics, but also in astrophysics and cosmology. One of the most well-known problems of SM is the problem of neutrino masses and neutrino oscillations \cite{osc_1_1, osc_1_2, osc_2_1, osc_2_2, osc_2_3}, which will be considered in this paper. The essense of the problem is the absence of clarity concerning what mechanism provides the observed neutrino mass spectrum and how neutrinos of one flavor disappear and turn into neutrinos of another flavor. Some still-unknown particles would be needed to analyse the neutrino puzzle. Let us mention that the solution of other equally well-known problems of SM, which are the problem of dark matter and dark energy in the Universe, as well as the problem of its baryon asymmetry (BAU) generation, like the neutrino problem, can also be associated with the {SM extension} by new particles and/or interactions. The inability of modern collider experiments to detect new particles may be due not only to their large masses, but also to extremely elusive interactions with SM particles, making their production extremely rare events. In other words, to detect feebly interacting particles, the requirements for the integrated luminosity of the beam(s) in the experiment must cross the 'intensity frontier'. 

A well-known example of a field-theoretic model, within the framework of which it is possible to solve a significant part of the above-mentioned problems by extending the set of SM particles with relatively light and weakly interacting new particles of the lepton sector, is the $\nu$MSM model (neutrino minimal standard model) \cite{nuMSM_1, nuMSM_2}. It is an extension of the SM with the help of three Majorana neutrinos whose flavor states do not participate in gauge interactions (so-called sterile neutrinos, also called heavy neutral leptons, HNL). The masses of sterile neutrinos do not exceed the electroweak scale ${\cal O} (10^2) $ GeV and at the same time significantly exceed the value of the mass parameter $Fv$ ({$F$} is the neutrino Yukawa coupling constant and $v$ is the Higgs field vacuum expectation value equal to 174~GeV), as a result of which the mass generation mechanism (known as the seesaw type-I \mbox{mechanism \cite{seesawI_1, seesawI_2, seesawI_3, seesawI_4},} see Section~\ref{sec:model} below) works well for neutrino. Within the framework of the $\nu$MSM, it is not necessary to introduce a new mass scale for heavy fermions, the Majorana masses are comparable to the masses of light quarks and charged leptons. The number of sterile neutrinos should be at least three. At least two states of sterile neutrinos are necessary to explain the known experimental parameters $\Delta m^2_{atm}$ and $\Delta m^2_{sol}$. However, none of these states are suitable for the role of a dark matter particle, since (1) their Yukawa constants are not sufficiently limited to form a moderate amount of dark matter within the framework of the Dodelson--Widrow mechanism \cite{dw}, and (2) the one-loop radiative decays of such states in the subdominant channel  
$N_{2,3} \to \nu \gamma$ with the energy $E_\gamma = 1/2 M_{2,3}$ \cite{7keV_th_1, 7keV_th_2} would produce too many gamma rays compared to the available astrophysical experimental data\footnote{
Note that the unidentified spectral line at the energy $E \sim 3.5$ keV in the stacked X-ray spectra of dark-matter-dominated objects has been observed \cite{7keV_exp_1, 7keV_exp_2}. The signal distribution over the sky and the ratios of its strengths are consistent with predictions for decaying dark matter with the mass $M_1 \simeq 7.1 \pm 0.1$ keV \cite{alehin}.}.
For these reasons, one more sterile neutrino (the HNL of the first generation\footnote{
Throughout the paper we will assume that there is a normal ordering in the HNL sector, i.e., $M_1 < M_2 \leq M_3$.}
added to the $\nu$MSM provides a suitable candidate for the role of dark matter. 

 Adding three HNLs introduces 18~new parameters, 3 Majorana masses and 15 parameters of the Yukawa coupling matrix $F$.
The mass of a DM sterile neutrino can be varied in the range 
$\sim$1--50 keV in the case of $\nu$MSM \cite{alehin, 1-50keV}. The baryogenesis of the $\nu$MSM is provided by the oscillation mechanism~\cite{alehin}. In the following, three generations of HNL are considered, which makes it possible to explain simultaneously the origin of the masses of standard (or active) neutrinos due to the seesaw mechanism, the existence of a dark-matter particle and the generation of baryon asymmetry. 
Significant limitations on the mixing parameter value for HNL in the $\nu$MSM model, as well as on the second- and third-generation HNL masses, which should be split very weakly \cite{2007_Shap_sym}, are reviewed in \cite{alehin}.

In extensions of the SM lepton sector, the active neutrino flavor states and mass
states can be related with the help of Pontecorvo--Maki--Nakagawa--Sakata mixing matrix 
$U_{\rm PMNS}$,
$\nu_{l L}(x) = \sum_{i} (U_{\rm PMNS})_{li} \upnu_{i L}(x)$,
where $\nu_{l L}$ is the left-handed (LH) component of the flavour
field, $\upnu_{i L}$ is the LH component of the state possessing a mass $m_i$,  $i=1,2,3$, and $l=e, \mu, \tau$~\cite{osc_1_1, osc_1_2, osc_2_1, osc_2_2, osc_2_3, pmns_3}. 
The exact values of neutrino masses are unknown, but it follows
from the 
existing data 
that there are at least three active neutrino mass eigenstates, 
neutrinos must be light, 
$|(m_\nu)_{ll'}| \leq 1$ eV ($l,l'=e, \mu, \tau$),
\cite{merle, 0nubb} and the neutrino mass values are rather strongly split.
The differences between them measured experimentally give the central values of mass parameters squared
$\Delta m_{sol}^2 \approx 7.54 \times  10^{-5}$ eV$^2$ and  
$\Delta m_{atm}^2 \approx 2.43 \times  10^{-3}$ eV$^2$  \cite{GA}. This leaves two different possibilities of neutrino mass ordering: normal ordering (NO) $m_1 < m_2 \ll m_3$ or inverse ordering (IO)  $m_3 \ll m_1 < m_2$. 
The mass of the lightest
neutrino $m_0$ may vary from zero to 
0.03 eV (NO) or 0.015 eV (IO) allowed by cosmological observations \cite{Plank} (see also \cite{drewes_22}). Within the $\nu$MSM,  the lightest active neutrino is lighter than ${\cal O} (10^{-5})$ eV \cite{nuMSM_1, nuMSM_2, nuMSM_lightest}. The HNL masses responsible for active neutrino mass generation
are rather strongly constrained both by direct searches 
\cite{MN<1eV_1, MN<1eV_2, HNLser_dir1, HNLser_dir2}
and cosmological considerations \cite{HNLser_cosm1, HNLser_cosm2, HNLser_cosm3}.
The analysis of lepton universality within the $\nu$MSM shows that
$M_{2,3} > 173$ MeV (264 MeV) for NO (IO)~\cite{lep_univ}, see also \cite{MN>100}. 
The recent investigation of BAU \cite{canetti_13} implies the masses of HNL to be $M_{2,3} \geq {\cal O}(3)$ GeV.

Historically, the motivation for considering right-handed (RH)
neutrinos have appeared from the left-right symmetric extension of the SM \cite{mp}
where the electroweak group of symmetry $SU(2)_L \times U(1)_Y$ was 
extended to the $SU(2)_L \times SU (2)_R \times U(1)_{B-L}$.
The observed left chirality of the low-energy weak interactions in the model with left--right symmetry, usually attributed to the spontaneous breaking of this symmetry, implies the existence of a right-handed neutrino.

It is commonly assumed that the effective mass $(m_{\nu})_{ll'}$ can be presented as
\begin{equation}
(m_\nu)_{ll'} = - M_D M_M^{-1} M_D^T 
\label{mnu_com}
\end{equation}
in the limit $M_D \ll M_M$, where $M_D, M_M$ are Dirac and Majorana mass matrices of the LH and RH neutrinos (see details below).
However, Eq. (\ref{mnu_com}) is a standard approximation of a more complicated expression including an additional $3 \times 3$ matrix $\theta$ ($|\theta| \ll I$) describing mixing of the active neutrinos with HNL. The contribution of the next-order terms in effective neutrino mass representations requires a separate study.  
In the following, Section \ref{sec:model}, we investigate the standard approach for the
type-I seesaw mechanism and consider the non-minimal approximations for effective masses of the active and sterile neutrinos. Nonstandard parametrization of the mixing inspired by diagonalization of the active-sterile neutrino mass matrix in the approach of \cite{casas_ibarra} and a nontrivial relation for the parametrization matrix $\Omega_{\rm nm}$ have been obtained. The numerical analysis within the $\nu$MSM framework is performed in Section \ref{sec:num}. Results are discussed in Section \ref{sec:concl}. 
A symmetric structure of effective and Majorana mass matrices as well as relations for $M_D \theta^T$ and $\theta^\dagger M_D$ are analyzed in Appendix \ref{sec:rel-MDR}.
Simple representations for the matrix $\Omega_{\rm nm}$ are considered in Appendix~\ref{sec:Omn_gen}.
A form of effective mass matrix of active neutrinos $m_\nu$ as function of $U_{\rm PMNS}$ in standard and non-minimal approximations is derived in Appendix \ref{sec:mnu}.

\section{A model with three right-handed neutrinos}
\label{sec:model}

The lagrangian of the SM extension by three right-handed neutrinos $\nu_R$ has the form (see e.g.,  \cite{white})
\begin{equation}
{\cal L} = {\cal L}_{SM} + i \overline{\nu}_R \partial_\mu \gamma^\mu \nu_R- \left( F \overline{l}_L \nu_R \tilde{H} + \frac{M_M}{2} \overline{\nu^c}_R \nu_R + \text{h.c.} \right),
\label{lagr}
\end{equation}
where 
$
l_L=(\nu_L \, \, e_L)^T
$
is the left-handed lepton doublet,
$\nu_R$ is the right-handed neutrino 
[$(\nu_R)^c \equiv C \overline{\nu}_R^T$, where $C=i \gamma_2 \gamma_0$ is the charge conjugation matrix], 
$H$ is the Higgs doublet ($\tilde{H}= \epsilon_{ij} H^\dagger$, $\epsilon_{ij}$ antisymmetric tensor), 
$F$ is the neutrino Yukawa coupling matrix, and
$M_M$ is the $3 \times 3$ Majorana mass matrix of the RH neutrinos,
which does not break the SM gauge symmetry.

After spontaneous electroweak symmetry breaking, the full neutrino mass term in Equation~(\ref{lagr}) can be presented as  
\begin{equation}
\frac{1}{2} (\overline{\nu}_L \vspace{2mm} \overline{\nu^c}_R) {\cal M} \left(
\begin{tabular}{c}
$\nu_L^c$\\
$\nu_R$ 
\end{tabular}
\right), \qquad
\text{where} \qquad
 {\cal M}=
\left(
\begin{tabular}{cc}
 0 & $M_D$\\
 $M_D^T$ & $M_M$  
\end{tabular}
\right),
\label{M66}
\end{equation}
$M_{D}= F v$ is a $3 \times 3$ neutrino Dirac mass matrix ($v$=174 GeV is a vacuum expectation value of the Higgs doublet), and $M_D, M_M$ are complex-valued in the general case.  
In the neutrino mass basis with the help of a unitary matrix ${\cal U}$
\begin{equation}
{\cal U}= W \cdot {\rm diag}(U_\nu,U_N^*),
\end{equation}
($U_\nu,U_N$ on the one hand side and $W$ on the other hand side are $3\times 3$ and $6\times 6$ matrices, respectively), one can define the diagonal mass matrix \cite{white}
\begin{equation}
{\cal U}^\dagger {\cal M} {\cal U}^*
=
\left(
\begin{tabular}{cc}
 $\hat{m}$ & 0 \\
 0 & $\hat{M}$  
\end{tabular}
\right),
\qquad
\begin{tabular}{c}
$\hat{m} = {\rm diag}(m_1, m_2, m_3)$,\\
$\hat{M} = {\rm diag}(M_1, M_2, M_3)$, 
\end{tabular}
\end{equation}
where $m_i, M_i$ are masses of light and heavy neutrinos.
The unitary diagonalization matrix $W$ formally expressed as the exponential of an antihermitian matrix, see \cite{casas_ibarra},
\begin{equation}
W = \exp \left(
\begin{tabular}{cc}
0 & $\theta$ \\
$-\theta^\dagger$ & 0
\end{tabular}
\right)
\label{W}
\end{equation}
(where $\theta$ is a $3 \times 3$ complex matrix)
converts the matrix ${\cal M}$ to a block diagonal form
\begin{equation}
 W^\dagger \left(
\begin{tabular}{cc}
 0 & $M_D$\\
 $M_D^T$ & $M_M$  
\end{tabular}
\right) W^* 
=
\left(
\begin{tabular}{cc}
$m_\nu$ & 0\\
0 & $M_N$ 
\end{tabular}
\right) \equiv \tilde{{\cal M}}.
\label{Mtilde} 
\end{equation}
Here, $m_\nu, M_N$ are effective Majorana mass terms for the LH and RH flavour neutrinos (see, e.g., \cite{bilenky, molinaro}) which are related with $\hat{m}$ and $\hat{M}$ by 
\begin{eqnarray}
m_\nu = U_\nu \hat{m} U_\nu^T,\ \qquad
M_N = U_N^* \hat{M} U_N^\dagger. 
\label{eff-hat}
\end{eqnarray}
%

The flavour states $\nu_L$ and $\nu_R$ and the mass states $\upnu$ and $N$ then can be related as \cite{white}
\begin{equation}
\left(
\begin{tabular}{c}
$\nu_L$\\
$\nu_R^c$ 
\end{tabular}
\right)
=
P_L {\cal U}
\left(
\begin{tabular}{c}
$ {\upnu} $\\
$N$ 
\end{tabular}
\right),
\label{states}
\end{equation}
where 
$P_L$ is {a} left-helicity projector.

\subsection{Standard approximation} 
Assuming that $\theta$ is “small,” it is sufficient to expand the rotation matrix (\ref{W}) up to ${\cal O}(\theta^2)$~terms
\begin{eqnarray}
W = \exp \left(
\begin{tabular}{cc}
0 & $\theta$ \\
$-\theta^\dagger$ & 0
\end{tabular}
\right)
\simeq
\left(
\begin{tabular}{cc}
$1-\frac{1}{2} \theta \theta^\dagger $ & $\theta $\\
$-\theta^\dagger$ & $1-\frac{1}{2} \theta^\dagger \theta$ 
\end{tabular}
\right), 
\label{W-lit}
\end{eqnarray}
thus, using (\ref{Mtilde}) and the requirement $\tilde{\cal M}_{12}=0$, one can relate Dirac, Majorana and $\theta$ matrices 
\begin{equation}
\theta \simeq M_D M_M^{-1}, \label{M_M-lit}
\end{equation}
which yields the expressions
\begin{eqnarray}
m_\nu & \simeq & - M_D \theta^T \simeq 
 - M_D M^{-1}_M M_D^T, \label{mnu-lit} \\  
M_N  &\simeq &  
M_M +\frac{1}{2} (\theta^\dagger \theta M_M+M_M^T \theta^T \theta^*) \simeq M_M, \label{MN-lit}
\end{eqnarray}
where, in some cases, the terms ${\cal O}(\theta^2)$ in (\ref{MN-lit}) are omitted, so Majorana mass {matrices} before and after rotation coincide.
One can represent (\ref{MN-lit}) in terms of $\theta$ and $M_D$ using the relations (\ref{MDRT}) and (\ref{MDR*}) (see Appendix \ref{sec:rel-MDR})
\begin{equation}
M_N \simeq (\theta^{-1} +\theta^\dagger )M_D.
\label{MN_m}
\end{equation}

In the representation (\ref{W-lit}), the flavour states of active neutrinos can be expressed as [see (\ref{states})]
\begin{eqnarray}
\nu_L & \simeq & \left( 1- \frac{1}{2} \theta \theta^\dagger \right) U_\nu \upnu_L + \theta U_N^* N_L, \label{nuL-2} \\
\nu_R^c & \simeq & -\theta^\dagger U_\nu \upnu_L + \left(1-\frac{1}{2} \theta \theta^\dagger \right) U_N^* N_L,
\end{eqnarray}
where terms in front of $\upnu_L$ and $N_L$ have to be identified with PMNS and mixing matrices\footnote{
In general, the PMNS matrix can be presented as
$U_{\rm PMNS} = U_l^\dagger (1 + \eta)U_\nu$ \cite{molinaro}, where
$U_l$ diagonalizes the charged lepton mass matrix $m_l$, $U_l m_l m_l^\dagger U_l^\dagger= {\rm diag}(m_e^2, m_\mu^2, m_\tau^2)$, but throughout the paper, we assume that $U_l=I$.
}
\begin{equation}
U_{\rm PMNS} \simeq  (1+\eta) U_\nu, \qquad 
\Theta_{\rm m}  \simeq  \theta U_N^*. \label{Theta-m} 
\end{equation}
Here, the parameter $\eta=-1/2 (\theta \theta^\dagger)$ is considered as a deviation from the unitarity.

\subsection{Non-minimal approximation}
One can, however, consider the expansion of the matrix $W$, see Equation~(\ref{W}), up to ${\cal O}(\theta^3)$-terms
\begin{eqnarray}
W \simeq  
\left(
\begin{tabular}{cc}
$1-\frac{1}{2} \theta \theta^\dagger$ & $\theta - \frac{1}{6} \theta \theta^\dagger \theta$\\
$-\theta^\dagger + \frac{1}{6} \theta^\dagger \theta \theta^\dagger$ & $1-\frac{1}{2} \theta^\dagger \theta$ 
\end{tabular}
\right), \label{W3}
\end{eqnarray}
where again, {$|\theta| \ll I$}.
Then, the corresponding expressions (\ref{Mtilde}) can be presented as
\begin{eqnarray}
\tilde{\cal M}_{11} & = & - (\theta M_D^T+M_D \theta^T) + \theta M_M \theta^T + \frac{1}{6} (\theta \theta^\dagger \theta M_D^T+ M_D \theta^T \theta^* \theta^T) \nonumber \\
&+& \frac{1}{2} (\theta \theta^\dagger M_D \theta^T+\theta M_D^T \theta^* \theta^T) 
\simeq m_\nu, \label{tilde11} \\
\tilde{\cal M}_{22} & = & M_M+ (\theta^\dagger M_D+ M_D^T \theta^*)- 
\frac{1}{2} (\theta^\dagger \theta M_M + M_M \theta^T \theta^*) \nonumber \\
&-& \frac{1}{2} \theta^\dagger (\theta M_D^T+M_D \theta^T) \theta^* 
- \frac{1}{6} (\theta^\dagger \theta \theta^\dagger M_D+ M_D^T \theta^* \theta^T \theta^*)
 \simeq  M_N, \label{tilde22} \\
\tilde{\cal M}_{12} & = & M_D - \theta M_M -\frac{1}{2}(\theta \theta^\dagger M_D+M_D \theta^T \theta^*) \nonumber \\
&-& \theta M_D^T \theta^*  + \frac{1}{2} \theta M_M \theta^T \theta^* + \frac{1}{6} \theta \theta^\dagger \theta M_M
\simeq 0. \label{tilde12}
\end{eqnarray} 

One can note that the relation between Dirac, Majorana, and $\theta$ matrices can be defined in the form [see (\ref{tilde12})]
\begin{equation}
M_M - \frac{1}{2} M_M \theta^T \theta^*-\frac{1}{6} \theta^\dagger \theta M_M 
 \simeq \theta^{-1} M_D \left(1-\frac{1}{2}\theta^T \theta^*\right)-\frac{1}{2}\theta^\dagger M_D - M_D^T \theta^*, \label{MM-MDR}
\end{equation}
which is more complicated than the expression (\ref{M_M-lit}) of the standard approximation
and can be converted to it 
if ${\cal O}(\theta^2)$-terms are neglected. 
As one can see, the relation between two mass scales $M_D$, $M_M$ and $\theta$ is not so obvious, but analyzing the relation (\ref{MM-MDR}) one can note that (see Appendix \ref{sec:rel-MDR})
\begin{equation}
{\cal O}(\theta^2 M_M) \sim  {\cal O}(\theta M_D). 
\label{rel}
\end{equation}
This allows one to express Equations (\ref{tilde11}) and (\ref{tilde22}) in terms of ${\cal O}(\theta^n M_D)$, where $n=-1,1,2,3$. However, the ${\cal O}(\theta^2 M_D)$ and ${\cal O}(\theta^3 M_D)$ terms are much smaller than ${\cal O}(\theta M_D)$, so in the following we will consider only the last one. Thus,  the expressions for $m_\nu, M_N$ and $M_M$ take the forms
\begin{eqnarray}
m_\nu &\simeq & -\theta M_D^T \simeq -M_D \theta^T, \label{mnu}\\
M_N &\simeq & 
\left( \theta^{-1} -\frac{1}{3} \theta^\dagger \right) M_D 
 \simeq  M_M + \theta^\dagger M_D, \label{MN}
\end{eqnarray}
where the right forms can be obtained if one takes into account the relations (\ref{MDRT}) and (\ref{MDR*}), see Appendix \ref{sec:rel-MDR}.

Analyzing the representation for the flavour states (\ref{states})
\begin{eqnarray}
\nu_L & \simeq & \left( 1- \frac{1}{2} \theta \theta^\dagger \right) U_\nu \upnu_L + \left( \theta -\frac{1}{6} \theta \theta^\dagger \theta \right) U_N^* N_L, \label{nuL-3} \\
\nu_R^c & \simeq & \left(-\theta^\dagger+\frac{1}{6} \theta^\dagger \theta \theta^\dagger \right) U_\nu \upnu_L + \left(1-\frac{1}{2} \theta^\dagger \theta \right) U_N^* N_L,
\end{eqnarray}
one can identify the PMNS and active-sterile mixing matrices with
\begin{equation}
U_{\rm PMNS} \simeq \left(1 - \frac{1}{2} \theta \theta^\dagger \right) U_\nu,
\label{Upmns-nm}
\qquad
\Theta_{\rm nm} \simeq \left( \theta -\frac{1}{6} \theta \theta^\dagger \theta \right) U_N^*. 
\end{equation}
As far as {$|\theta| \ll 1$}, one can omit the terms ${\cal O}(\theta^n)$, $n>1$ in the expressions (\ref{Upmns-nm}), so these forms coincide with ones of the standard approximation, see (\ref{Theta-m}).

Using (\ref{MN_m}) and (\ref{MN}), one can obtain 
the difference between the effective mass of heavy neutrinos in standard and non-minimal approximations 
\begin{equation}
M_{N {\rm m}} - M_{N {\rm nm}} = \frac{4}{3} \theta^\dagger F v.
\end{equation}

\subsection{Parametrization of the mixing matrix $\Theta$}

\subsubsection{Standard approximation} 

The expression $I = M_N M_N^{-1}$, with the help of (\ref{eff-hat}), (\ref{mnu-lit}), (\ref{MN-lit}) and the assumption $U_{\rm PMNS} \simeq U_\nu$ can be presented as
\begin{eqnarray}
I 
= -M_D^T U_{\rm PMNS}^* \hat{m}^{-1} U_{\rm PMNS}^\dagger M_D U_N \hat{M}^{-1} U_N^T.
\label{M_Nunit}
\end{eqnarray}
One can rewrite (\ref{M_Nunit}) in the form 
\begin{equation}
I= - (\sqrt{\hat{M}^{-1}} U_N^T M_D^T U_{\rm PMNS}^* \sqrt{\hat{m}^{-1}})(\sqrt{\hat{m}^{-1}}U_{\rm PMNS}^\dagger M_D U_N \sqrt{\hat{M}^{-1}})
= \Omega_{\rm m}^T \Omega_{\rm m},
\end{equation}
where 
\begin{equation}
\Omega_{\rm m} = i \sqrt{\hat{m}^{-1}}U_{\rm PMNS}^\dagger M_D U_N \sqrt{\hat{M}^{-1}}
\label{Omega_m}
\end{equation}
is an arbitrary complex orthogonal matrix ($\Omega_{\rm m}^T \Omega_{\rm m}=I$). 
Thus, the Dirac matrix can be parametrized as \cite{casas_ibarra}
\begin{equation}
M_D = - i U_{\rm PMNS} \sqrt{\hat{m}} \Omega_{\rm m} \sqrt{\hat{M}} U_N^\dagger,
\label{cas-ib-par}
\end{equation}
which together with (\ref{M_M-lit}) leads to the expression for the  mixing matrix
\begin{equation}
\label{Theta-CI}
\Theta_{\rm m} \simeq \theta U_N^*  \simeq  - i U_{\rm PMNS} \sqrt{\hat{m}} \Omega_{\rm m} \sqrt{\hat{M}^{-1}}.
\end{equation}

\subsubsection{Non-minimal approximation}

The same procedure $I = M_N M_N^{-1} $ in the non-minimal case $M_M \sim {\cal O}(\theta M_D)$ leads to
\begin{eqnarray}
- \left[ \sqrt{\hat{M}^{-1}} U_N^T \left(\theta^{-1}-\frac{1}{3}\theta^\dagger \right) U_\nu \sqrt{\hat{m}} \right] \left[ \sqrt{\hat{m}}  U_\nu^T (\theta^T)^{-1} U_N \sqrt{\hat{M}^{-1}} \right] \simeq I,
\label{I}
\end{eqnarray}
where $M_N^{-1} \equiv U_N \hat{M}^{-1} U_N^T$ and  
$M_N$ is defined by Equations~(\ref{MN}) and (\ref{mnu}). 
If one denotes the expression in the right brackets of (\ref{I}) as $\Omega_{\rm nm}$
\begin{equation}
\Omega_{\rm nm} = i \sqrt{\hat{m}} U_\nu^T (\theta^T)^{-1} U_N \sqrt{\hat{M}^{-1}}
\label{Omega-nm}
\end{equation}
and takes into account that the left side of (\ref{I}) concerns the unit matrix, then one can conclude using the identity $\Omega_{\rm nm}^{-1} \Omega_{\rm nm}=I$ that the expression in left brackets is $\Omega_{\rm nm}^{-1}$. 
Then, from Equation~(\ref{I}) one can find the relation for $\Omega_{\rm nm}$
\begin{equation}
\Omega_{\rm nm}^{-1} = \Omega_{\rm nm}^T+\frac{1}{3} \hat{M}^{-1} (\Omega_{\rm nm}^{-1})^* \hat{m}.  
\label{Omega-nm-relation} 
\end{equation}

Defining $\theta$ with the help of (\ref{Omega-nm}) 
\begin{equation}
\theta \simeq -i U_{\nu} \sqrt{\hat{m}} (\Omega_{\rm nm}^{-1})^T \sqrt{\hat{M}^{-1}} U_N^T
\label{theta_nm}
\end{equation}
and using the definition of the mixing matrix $\Theta_{\rm}$, Equation~(\ref{Theta-m}), one can write
\begin{equation}
\Theta_{\rm nm} \simeq \theta U_N^* \simeq 
-i U_{\rm PMNS} \sqrt{\hat{m}} (\Omega_{\rm nm}^{-1})^T \sqrt{\hat{M}^{-1}},
\label{Theta-RMD}
\end{equation}
where again it was assumed that $U_\nu \simeq U_{\rm PMNS}$.
Note that the mixing matrix $\Theta$ in approximations where $M_M \simeq \theta^{-1} M_D$ and $M_M \simeq {\cal O} (\theta M_D)$ is parametrized in the same way. The difference is in the presence of requirements that must be met by an arbitrary $3 \times 3$ matrix~$\Omega$. 
The corresponding Yukawa coupling matrix in terms of $\Omega_{\rm nm}$ can be presented as
\begin{equation}
F \simeq \frac{i}{v} U_{\rm PMNS} \sqrt{\hat{m}} \Omega_{\rm nm} \sqrt{\hat{M}} U_N^\dagger.
\label{F-RMD}
\end{equation}
Note that the Dirac matrix in non-minimal approximation differs from (\ref{cas-ib-par}) by the parametrization matrix $\Omega_{\rm nm}$ and the sign because sign($\Omega_{\rm nm}) \neq $ sign($\Omega_{\rm nm}^{-1}$) whereas sign($\Omega_{\rm m}$) = sign($\Omega_{\rm m}^{-1}$), see (\ref{Omega_m}) and (\ref{Omega-nm}). 

There is an infinite set of $\Omega$ matrices leading to the
observed neutrino parameters~\cite{molinaro}. 
In various mixing scenario frameworks, the matrix $\Omega_{\rm m}$ can be parametrized by means of three Euler angles (complex-valued in general), see \cite{DK_jetp, das_22}; one complex angle $\omega$ and the sign factor $\xi = \pm 1$ \cite{drewes_22, canetti_13}, or just by the unit matrix, see, e.g., \cite{DK_jetp, DFK}. 
It is desirable to relate a minimal and non-minimal matrices $\Omega_{\rm m}$ and $\Omega_{\rm nm}$ in order to consider the limit of the discussed approximations. However, one can find that the simple representation 
\begin{equation}
\Omega_{\rm nm} = \Omega_{\rm n} + \Omega_{\rm m}, 
\label{n+m}
\end{equation}
where $ \Omega_{\rm m}  \Omega_{\rm m}^T=I$, $ \Omega_{\rm n}  \Omega_{\rm n}^T \neq I$ is not relevant, see Appendix~\ref{sec:Omn_gen}.

\subsection{Interactions of the Majorana fields with SM particles}
\label{sec:currents}

The charged-current (CC) and the neutral-current (NC) interactions of the Majorana fields with SM particles in terms of mixing parameter and $U_{\rm PMNS}$ can be obtained as follows (see also \cite{DFK}). 
Using the SM normalization conventions of \cite{white}
\begin{equation}
- \frac{g}{\sqrt{2}} \overline{\nu}_L \gamma^\mu l_L W_\mu^+
- \frac{g}{\sqrt{2}} \overline{l}_L \gamma^\mu \nu_L W_\mu^-
-\frac{g}{2 \cos \theta_W} \overline{\nu}_L \gamma^\mu \nu_L Z_\mu,
\end{equation}
where $\theta_W$ is the Weinberg angle, one can write
the {corresponding neutral and charged currents} in the mass basis
\begin{eqnarray}
\label{nc_nu}
{\cal L}_{NC}^\upnu &=& 
\frac{g}{2 c_W} \overline{\upnu}_L \gamma^\mu U_{\rm PMNS}^\dagger U_{\rm PMNS} \upnu_L Z_\mu, \\
\label{cc-nu}
{\cal L}_{CC}^\upnu &=&
-\frac{g}{\sqrt{2}} \overline{l}_L \gamma^\mu U_{\rm PMNS} \upnu_L W_\mu^- + \text{h.c.}, \\
\label{nc-N}
{\cal L}_{NC}^N &=& 
-\frac{g}{2 c_W} \overline{N}_L \gamma^\mu \Theta^\dagger \Theta N_L Z_\mu - \left( \frac{g}{2 c_W} \overline{\upnu}_L U_{\rm PMNS}^\dagger \gamma^\mu \Theta N_L Z_\mu + h.c. \right), \\ 
\label{cc-N}
{\cal L}_{CC}^N &=& 
- \frac{g}{\sqrt{2}} \overline{l}_L \gamma^\mu \Theta N_L W_\mu^- + \text{h.c.} ,
\end{eqnarray}
where
$\Theta$ is a minimal $\Theta_{\rm m}$ or a non-minimal $\Theta_{\rm nm}$ mixing matrix.

It is instructive to note that the active neutrino mass matrix $m_\nu$ has the same representation for both $M_M \simeq \theta^{-1} M_D$ and $M_M \sim {\cal O}(\theta M_D)$ approximations
(see Appendix \ref{sec:mnu})
\begin{equation}
m_\nu  \simeq U_{\rm PMNS} \hat{m} U_{\rm PMNS}^T.
\label{mnu-same}
\end{equation}

\section{Numerical examples}
\label{sec:num}

As mentioned above, signals of sufficiently light HNL's are not observed in collider or beam-dump experiments. Experimental information is reduced to exclusion contours generated on the mass-mixing plane for different detectors. It should be noted that these exclusion contours use a model-independent phenomenological approach, in which only one HNL gives the observed signal, and the remaining two are strongly decoupled and practically do not affect the analysis. Within the framework of this approach, only two independent parameters are considered, the HNL mass and its coupling constant with an active neutrino, the generation of exclusion contours is carried out for the square of the mixing parameter as a function of mass for one flavor only, while the mixing for other flavors is assumed to be zero. Assumptions of this kind seem to be very strong and need to be translated for the case of a specific model. Suffice it to note that, even in the simple case of one generation of Majorana neutrinos \cite{grounau}, the mixing parameter $U=M_D/M_N$ depends on the mass of HNL, as a result of which the mapping of the active neutrino mass---HNL mass plane to the mixing squared---HNL mass plane is a triangle, not a rectangle. A part of the triangle for a limited region of HNL masses may look approximately like a rectangle  due to the low mass of the active neutrino.

In the following instructive example, we will vary the masses of the lightest neutrino and heavy neutrinos in  
the  ranges (see \cite{nuMSM_1, nuMSM_2, nuMSM_lightest})
\begin{eqnarray}
\label{m0scan}
m_0: && 0 - 10^{-5} \;\text{eV},\\ 
M_i: && 1 - 50 \;\text{keV} \;\text{for } M_1, \qquad 1-100 \;\text{GeV} \;\text{for } M_{2,3},
\label{Mscan}
\end{eqnarray}
parametrize the masses of active neutrinos as 
$
m_1^2 = m_0^2,
m_2^2 = m_0^2+ \Delta m_{sol}^2,
m_3^2 = m_0^2+ \Delta m_{sol}^2+  \Delta m_{atm}^2
$
for NO or 
$
m_1^2 = m_0^2+ \Delta m_{sol}^2,
m_2^2 = m_0^2+ \Delta m_{sol}^2+  \Delta m_{atm}^2,
m_3^2 = m_0^2
$
for IO 
and use the present central values of the known active neutrino sector parameters \cite{GA}.
In addition, we will use phenomenologically convenient quantities (see, e.g., \cite{drewes_22})
\begin{equation}
U_{\alpha I}^2= |\Theta_{\alpha I}|^2, \qquad
 U_I^2=\sum_\alpha U_{\alpha I}^2, \qquad
  U^2=\sum_I U_I^2.
\end{equation}

Among the infinite set of $\Omega_{\rm m}$ matrices satisfying the  requirement $\Omega_{\rm m}^T \Omega_{\rm m}=I$, 
we consider 
the following explicit forms 
\begin{eqnarray}
\Omega^{'} &=& I, \qquad
\Omega^{''}=
- \left(
\begin{tabular}{ccc}
$\frac{1}{\sqrt{2}}$ & $-\frac{1}{2}$ & $-\frac{1}{2}$ \\
$\frac{1}{3 \sqrt{2}}$ & $\frac{5}{6}$ & $-\frac{1}{2}$ \\
$\frac{2}{3}$ &  $\frac{1}{3 \sqrt{2}}$ & $\frac{1}{\sqrt{2}}$
\end{tabular}
\right), \\
\Omega^{'''}_{\rm NO}&=&
\left(
\begin{tabular}{ccc}
1 & 0 & 0\\
0 & 0 & 1 \\
0 & $-1$ & 0
\end{tabular}
\right),
\;\text{  } 
\Omega^{'''}_{\rm IO}=
\left(
\begin{tabular}{ccc}
0 & 0 & 1\\
0 & $-1$ & 0 \\
1 & 0 & 0
\end{tabular}
\right),
\label{Omega}
\end{eqnarray}
where $\Omega_{\rm m}=\Omega^{'}$ corresponds to a naive explicit form used, e.g., in \cite{DK_jetp, DFK};
the form 
$\Omega_{\rm m}=\Omega^{''}$ is not motivated by any special physical scenarios and used for numerical estimations in this section only; and
$\Omega_{\rm m}=\Omega^{'''}$ is a form inspired by a parametrization in \cite{canetti_13} with fixed parameters $\omega=\pi/2$, $\xi=1$
which (as was {determined} numerically) 
maximize the $CP$-asymmetry.
The non-minimal parametrization matrix satisfying (\ref{Omega-nm-relation}) is chosen as {(see, also, Appendix} \ref{sec:Omn_gen})
\begin{equation}
\Omega_{\rm nm} = {\rm diag} (\omega_1, \omega_2, \omega_3), \;\text{where }
\omega_i = \sqrt{1-\frac{1}{3} \frac{m_i}{M_i}}.
\label{Omega_nm}
\end{equation}

Results of evaluations for diagonal elements $(m_\nu)_{ee}$,  $(m_\nu)_{\mu \mu}$,  $(m_\nu)_{\tau \tau}$ with explicit forms $\Omega_{\rm m}$ as $\Omega^{'}, \Omega^{''}, \Omega^{'''}$ are presented in 
Table~\ref{tab:mnu}, where 
first $M_D$ is found using (\ref{cas-ib-par}) then $(m_\nu)_{ll}$ is evaluated with the help of (\ref{eff-hat}) and (\ref{mnu-lit}).
The results for all cases coincide [in accordance with (\ref{mnu-same})]
and are below the present experimental bounds.  
\begin{table} 
\begin{center}
\caption{Numerical estimations of diagonal elements $(m_\nu)_{ee}$,  $(m_\nu)_{\mu \mu}$ and  $(m_\nu)_{\tau \tau}$ in units of $10^{-2}$ eV for explicit forms $\Omega_{\rm m}$: $\Omega^{'}, \Omega^{''}, \Omega^{'''}$, see  (\ref{Omega}). Here, $m_0$ and $M_1, M_2, M_3$ are  defined by (\ref{m0scan}) and (\ref{Mscan}). \label{tab:mnu}}
\begin{tabular}{cccc}
\hline
\textbf{Ordering} & \textbf{$l=e$} & \textbf{$l=\mu$} & \textbf{$l=\tau$}\\
\hline
NO 			& $0.4 $ 	& $2.7 $ & $2.6$ \\ 
IO 		& $4.9 $ 	& $1.9 $ & $2.6 $ \\ 
\hline
\end{tabular}
\end{center}
\end{table}
\unskip

\begin{figure}
\begin{minipage}[h]{0.5\linewidth}
\center{\includegraphics[width=1\linewidth]{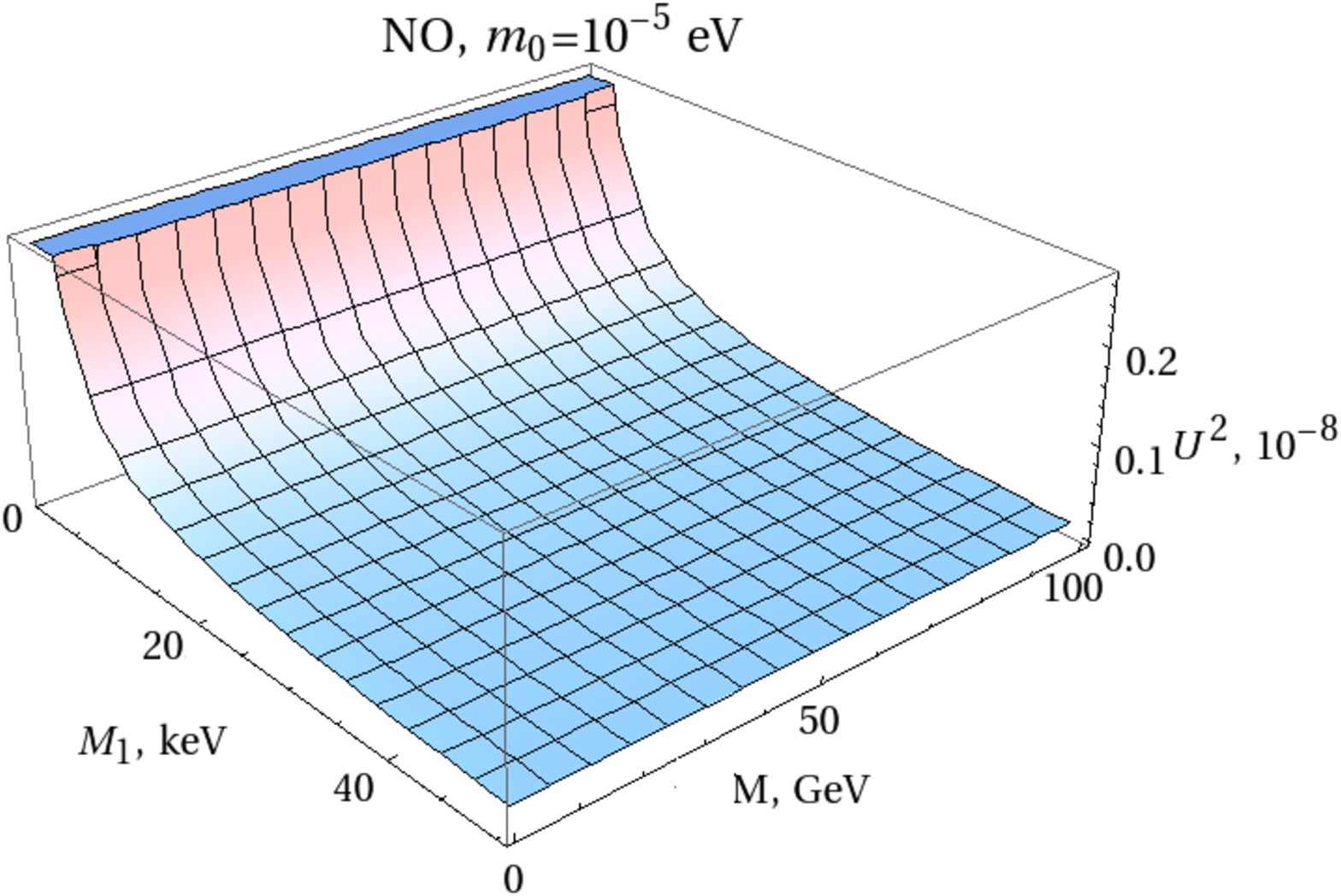}} \\ (a)
\end{minipage} 
\begin{minipage}[h]{0.5\linewidth}
\center{\includegraphics[width=1\linewidth]{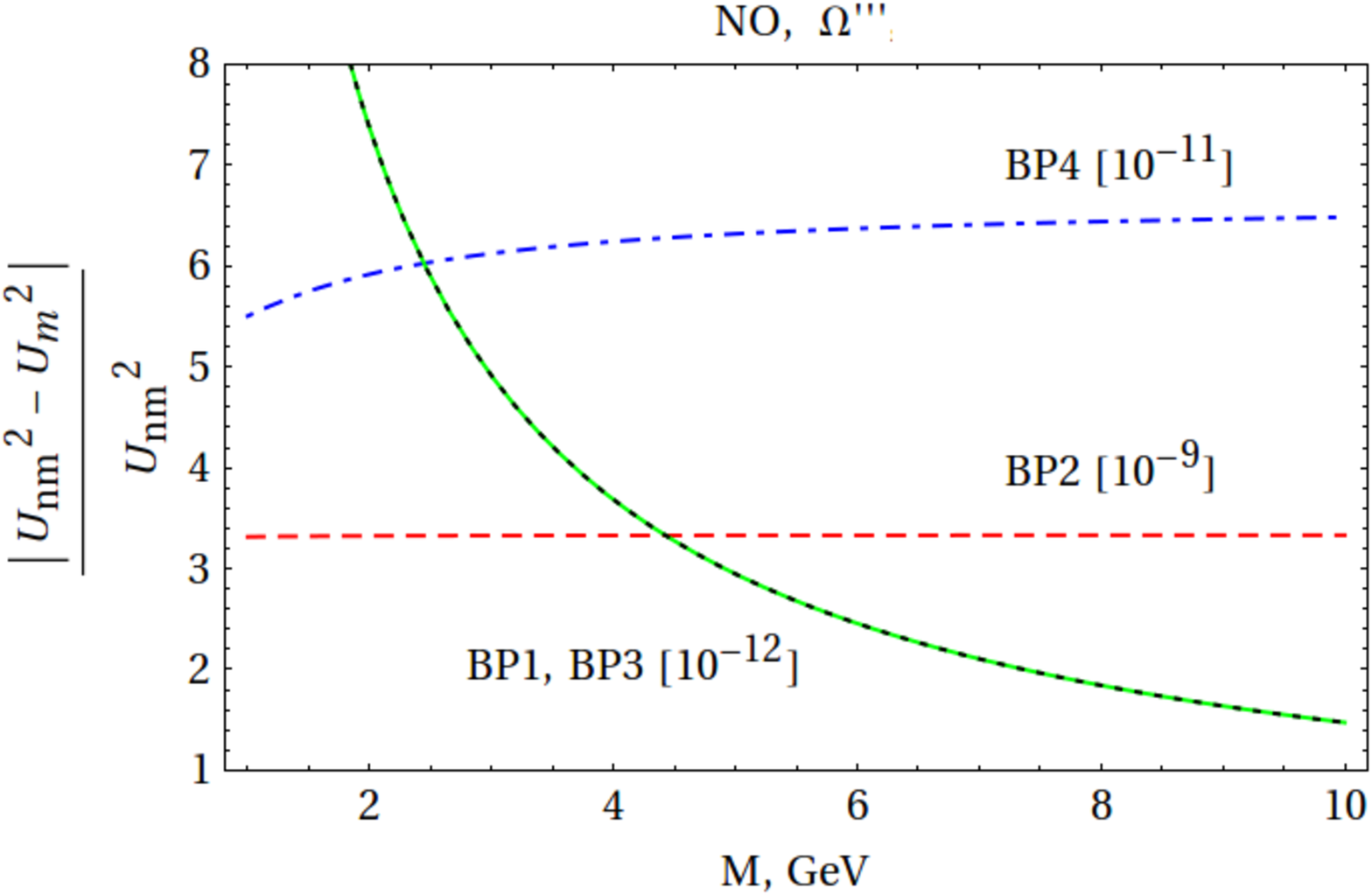}} \\ (b)
\end{minipage} 
\caption{Mixing parameter estimations for NO: 
(\textbf{a}) the dependence of $U^2$ as a function of masses of heavy neutrinos $M_1$ and $M_2=M_3=M$ at $m_0=10^{-5}$ eV, $\Omega_{\rm m}=\Omega^{'}$;
(\textbf{b}) the relative difference in $U^2$ obtained in minimal and non-minimal approximations as a function of $M_2=M_3=M$ at 
BP1: $m_0=0$, $M_1$=1 keV  in units of $10^{-12}$ (black dotted line);
BP2: $m_0=10^{-5}$ eV, $M_1$=1 keV  in units of $10^{-9}$ (red dashed line);
BP3: $m_0=0$, $M_1$=50 keV in units of $10^{-12}$ (green solid line);
and BP4: $m_0=10^{-5}$ eV, $M_1$=50 keV  in units of $10^{-11}$ (blue dot-dashed line). 
\label{fig1}}
\end{figure}

The dependence of $U^2$ as a function of masses of heavy neutrinos $M_1$ and $M_2=M_3=M$ at $m_0=10^{-5}$ eV 
for NO is presented in Figure~\ref{fig1}a, where one can see that its values are ${\cal O}(10^{-9})$ and practically constant for all  $M$. 
The qualitative behavior is similar for other cases. 
As a rule, the contributions of $M_1$ to $U^2$ are neglected (as was assumed e.g., in \cite{canetti_13});
however, Figure~\ref{fig1}a demonstrates that the dependence $U^2(M_1)$ can be sizable in some cases. 
One can note that the minimal (maximal) value of $U^2$ is achieved at benchmark point {BP$_{\rm min}$ (BP$_{\rm max}$)}, where
\vspace{7pt} \\
BP$_{\rm min}$:\, \, $m_0=0$, $M_1$=50 keV, $M$=100 GeV,\\
BP$_{\rm max}$:\, \, $m_0=10^{-5}$ eV, $M_1$=1 keV, $M$=1 GeV.
\vspace{7pt} 

Numerical estimations of $U^2$ for different $\Omega_{\rm m}$ ($\Omega^{'}, \Omega^{''}, \Omega^{'''}$) can be found in Table~\ref{tab:U2}{;} based on these numbers, it is possible to come to a conclusion about the great complexity of heavy Majorana neutrinos observation. 

\begin{table}
\begin{center}
\caption{Results for $U^2$ in the case of NO and IO for different $\Omega_{\rm m}$: $\Omega^{'}, \Omega^{''}, \Omega^{'''}$, see  (\ref{Omega}). Here, $m_0$ and  $M_1, M_2, M_3$ are defined by (\ref{m0scan}) and (\ref{Mscan}).\label{tab:U2}}
\begin{tabular}{cccc}
\hline
\textbf{Ordering} & $\Omega^{'}$ & $\Omega^{''}$ & $\Omega^{'''}$\\
\hline
NO 			& $\ 10^{-12} - 10^{-8} $ 	& $ 10^{-7} - 10^{-5} $ & $10^{-12} - 10^{-8}$ \\ 
IO 		& $ 10^{-6} - 10^{-5} $ 	& $ 10^{-6} - 10^{-5} $ & $10^{-12} - 10^{-8} $ \\ 
\hline
\end{tabular}
\end{center}
\end{table}
\unskip

The relative difference $|U^2_{\rm nm}-U^2_{\rm m}|/U^2_{\rm nm}$ for NO, IO and different $\Omega_{\rm m}$ ($\Omega^{'}, \Omega^{''}, \Omega^{'''}$) at  BP$_{\rm min}$ (BP$_{\rm max}$) is presented in Table~\ref{tab:deltaU2}. One can see that it is quite small for $\Omega^{'}, \Omega^{'''}_{\rm NO}$ and is very large for $\Omega^{''}_{\rm NO}$, which is a consequence of different representations for $\Omega_{\rm m}$ and $\Omega_{\rm nm}$. Thus, the most relevant comparison can be performed for $\Omega_{\rm m}=\Omega^{'}$ and $\Omega_{\rm nm}$ defined by (\ref{Omega}) and (\ref{Omega_nm}) since for such representations of the parametrization matrix, it is possible to trace the limit from the standard approximation to the non-minimal one.

\begin{table}
\begin{center}
\caption{
The relative difference $|U^2_{\rm nm}-U^2_{\rm m}|/U^2_{\rm nm}$  for NO, IO and different $\Omega_{\rm m}$ ($\Omega^{'}, \Omega^{''}, \Omega^{'''}$) (\ref{Omega}) at BP$_{\rm min}$ (BP$_{\rm max}$).\label{tab:deltaU2}}
\begin{tabular}{cccc}
\hline
\textbf{Ordering} & $\Omega^{'}$ & $\Omega^{''}$ & $\Omega^{'''}$\\
\hline
NO 			& $\ 10^{-13} \;\text{ } (10^{-9}) $ 	& $10^{6} \;\text{ } ( 10^{3})$ & $10^{-13} \;\text{ } ( 10^{-9})$ \\ 
IO 		& $ 10^{-7} \;\text{ } ( 10^{-5} )$ 	& $ 0.1 \;\text{ } (0.1)  $ &  $ 1 \;\text{ } (1) $ \\ 
\hline
\end{tabular}
\end{center}
\end{table}
\unskip

The relative difference of $U^2$ obtained in the minimal and non-minimal approximations as a function of $M_2=M_3=M$ for $\Omega^{'''}$ in the NO case is presented in Figure~\ref{fig1}b. 
 As one can see, the results for BP1 and BP3 coincide which is explained by the choice $m_0$=0 and the explicit form of $\Omega^{'''}$.
The behavior of the relative difference of $U^2$ at $m_0=0$ and $m_0 \neq 0$ is different due to the contributions of $M_1$ to the mixing parameter.  Thus, the approximation with two RH neutrinos is quite reliable at $m_0$=0 and $\Omega^{'''}$, while one should use a model with three  HNL at $m_0 \neq 0$. 
The relative difference for NO is rather small.
Thus, in the cases where the limit of discussed approximations can be traced, the phenomenological consequences with a mixing parameter obtained in
the non-minimal approximation can be hardly distinguished from the results of a standard approach (see examples in \cite{DK_jetp, DFK, DK_lep_un}) at present or future facilities.

\section{Conclusions}
\label{sec:concl}

The paper is devoted to the investigation of the SM extension by the three generations of sterile Majorana neutrinos which recover the left--right symmetry between quarks and leptons. 
Experimental detection of particles of this kind would greatly affect our understanding of particle physics and cosmology. The analysis carried out is especially significant for models in which there is no additional nonstandard scale of particle masses, so that a well-known 'model-independent approach' to calculating constraints for the model parameter space may not be directly relevant to reality. 
A non-minimal approximation for effective masses of light and heavy neutrinos in the framework of a type-I seesaw mechanism with three generations of right-handed Majorana neutrinos was considered. It was found that the obtained next-order corrections to the effective mass matrix of HNL can influence the well-known results of the current order modifying them with ${\cal O}(\theta M_D)$ corrections. 
Additional contributions to the effective masses of neutrinos were analysed and the regimes when they can be significant were defined.  
Following the method proposed in~\cite{casas_ibarra}, a parametrization of the mixing matrix was performed with the help of a complex $3 \times 3$ matrix $\Omega_{\rm nm}$ satisfying a nontrivial relation (\ref{Omega-nm-relation}) in the non-minimal approximation. 
The structure of effective and Majorana mass matrices was scrutinised and it was demonstrated that they are symmetric matrices, a fact which has not been discussed before. 
Useful relations were found for  $M_D \theta^T$ and $\theta^\dagger M_D$, which allow one to express effective masses in a simple form. 
It is interesting to note that the modification of the mixing matrix does not change results for the effective mass matrix of active neutrinos $m_\nu$ {in the non-minimal case}. 

The obtained results were analyzed in detail within the $\nu$MSM framework.
It was demonstrated that the approximation with two RH neutrinos is quite reliable at $m_0$=0, while one should use a model with three  HNL at $m_0 \neq 0$.
 New additional corrections to effective masses can be more pronounced when heavy neutrino masses are small and the lightest mass of active neutrino is of ${\cal O}(10^{-5})$ eV.
Numerical examples demonstrate that the relative difference in the mixing parameter obtained in  standard and non-minimal approximations is negligibly small for all benchmark scenarios of the model where the limit of the discussed approximations can be traced. 
Thus, phenomenological consequences for the mixing parameter obtained in the minimal and non-minimal approximations can be hardly distinguished at current and future facilities.

The question of the explicit form of the parametrization matrix $\Omega$ in standard and non-minimal approximations is still ambiguous. Various forms satisfying the requirement {$|\theta| \ll I$} {and $\Omega_{\rm m} \Omega_{\rm m}^T=I$} were considered with significantly different results. 
Effective masses of HNL in the standard and non-minimal approximations demonstrate {noticeable}  differences due to new sizeable terms of effective heavy neutrino mass matrix.
Note that ${\cal O}(\theta M_D)$ corrections to the splitting of eigenvalues of $M_M$ and $M_N$ matrices can impact both the lepton-number-violating event rates at colliders \cite{drewes_LNV} and 
leptogenesis in the early Universe \cite{shap_leptogenesis} (see also \cite{drewes_22}).

\vspace{24pt} 

The research was carried out within the framework of the scientific program of the National Center for Physics and Mathematics, project ''Particle Physics and Cosmology''.

\appendix
\section{Symmetric matrices \boldmath{$m_\nu, M_N, M_M$} and relations for \boldmath{$M_D \theta^T$} and \boldmath{$\theta^\dagger M_D$}}
\label{sec:rel-MDR}

The elements of the matrix $\tilde{{\cal M}}$, see (\ref{Mtilde}), have the forms 
\begin{eqnarray}
\tilde{{\cal M}}_{11} &=& W_{11}^\dagger M_D W_{21}^*+W_{21}^\dagger (M_D^T W_{11}^*+M_M W_{21}^*) \equiv m_\nu \label{11}\\
\tilde{{\cal M}}_{22} &=& W_{12}^\dagger M_D W_{22}^*+W_{22}^\dagger (M_D^T W_{12}^*+M_M W_{22}^*) \equiv M_N, \label{22}\\
\tilde{{\cal M}}_{12} &=& W_{11}^\dagger M_D W_{22}^*+W_{21}^\dagger (M_D^T W_{12}^*+M_M W_{22}^*) \equiv 0, \label{12}\\
\tilde{{\cal M}}_{21} &=& W_{12}^\dagger M_D W_{21}^*+W_{22}^\dagger (M_D^T W_{11}^*+M_M W_{21}^*) \equiv 0. \label{21}
\end{eqnarray}

Analysing (\ref{11})--(\ref{21}), one can note that $m_\nu, M_N$ and $M_M$ are symmetric matrices
\begin{equation}
m_\nu = m_\nu^T, \, \,
M_N = M_N^T, \, \,
M_M = M_M^T,
\label{x=xT}
\end{equation}
where for the last expression the relaton $\tilde{\cal M}_{12}-\tilde{\cal M}_{21}^T=0$ was used. 

In the \textit{standard approximation}, a requirement $\tilde{\cal M}_{12}=0$ [see (\ref{Mtilde})] has a form
\begin{equation}
\theta M_M \simeq M_D - \frac{1}{2} (\theta \theta^\dagger M_D+M_D \theta^T \theta^*)-\theta M_D^T \theta^*.
\label{12min}
\end{equation}

Multiplying (\ref{12min}) by $\theta^T$ from the right side
\begin{equation}
\theta M_M \theta^T \simeq M_D \theta^T - \frac{1}{2} (\theta \theta^\dagger M_D \theta^T +M_D \theta^T \theta^* \theta^T)-\theta M_D^T \theta^* \theta^T
\end{equation}
and omitting terms ${\cal O}(\theta^3)$ (as far as $|\theta| \ll I$), one can find 
\begin{equation}
M_D \theta^T \simeq \theta M_D^T, \label{MDRT}
\end{equation}
where the relation $\theta M_M \theta^T = (\theta M_M \theta^T)^T$ [see (\ref{x=xT})] was applied. 
The same relation (\ref{MDRT}) takes place for the \textit{non-minimal approximation}, where instead of (\ref{12min}) one should use Equation~(\ref{MM-MDR}). 

Multiplying (\ref{12min}) by $\theta^{-1}$ from the left side
\begin{equation}
M_M \simeq \theta^{-1} M_D - \frac{1}{2} (\theta^\dagger M_D  +\theta^{-1} M_D \theta^T \theta^*)- M_D^T \theta^* 
\end{equation}
and using the relation (\ref{MDRT}), one can rewrite the requirement $M_M = M_M^T$ in the form
\begin{equation}
\theta^\dagger M_D \simeq M_D^T \theta^*.
\label{MDR*}
\end{equation}

The relation (\ref{MDR*}) is also relevant in the \textit{non-minimal approximation}. The expression (\ref{MM-MDR}) can be written as
\begin{eqnarray}
\theta^\dagger \theta M_M &-& \frac{1}{2} \theta^\dagger \theta M_M \theta^T \theta^* -\frac{1}{6} \theta^\dagger \theta \theta^\dagger \theta M_M \simeq \nonumber \\
&& \theta^\dagger M_D-\frac{1}{2} \theta^\dagger \theta \theta^\dagger M_D-\frac{1}{2} \theta^\dagger M_D \theta^T \theta^*-\theta^\dagger \theta M_D^T \theta^*,
\end{eqnarray}
so, in the limit ${\cal O}(\theta M_D)$ 
\begin{equation}
\theta^\dagger \theta M_M \simeq \theta^\dagger M_D, \qquad
M_M \theta^T \theta^* \simeq M_D^T \theta^*,
\end{equation}
which allows one to express (\ref{MM-MDR}) as
\begin{equation}
M_M \simeq \theta^{-1} M_D -\frac{1}{3} \theta^\dagger M_D -\frac{1}{2} M_D^T \theta^* - \frac{1}{2} \theta^{-1} M_D \theta^T \theta^*.
\end{equation}

The requirement $M_M=M_M^T$, (\ref{x=xT}), with the help of (\ref{MDRT}), leads to
\begin{equation}
\theta^{-1} M_D+\frac{2}{3} \theta^\dagger M_D \simeq M_D^T (\theta^{-1})^T+\frac{2}{3} M_D^T R^*.
\label{nm-eq}
\end{equation}

Multiplying (\ref{nm-eq}) by $\theta$ and $\theta^T$ from the left and right sides, one can obtain
\begin{equation}
\theta \theta^\dagger M_D \theta^T \simeq \theta M_D^T \theta^* \theta^T,
\end{equation}
which leads to the relation (\ref{MDR*}).

\section{A simple representation for the matrix \boldmath{$\Omega_{\rm nm}$}}
\label{sec:Omn_gen}

Suppose that
\begin{equation}
\Omega_{\rm nm} = \Omega_{\rm n}+\Omega_{\rm m}, 
\label{n+m}
\end{equation}
where $ \Omega_{\rm m}  \Omega_{\rm m}^T=I$, $ \Omega_{\rm n}  \Omega_{\rm n}^T \neq I$. Inserting (\ref{n+m}) into (\ref{Omega-nm-relation}), one can find 
\begin{equation}
\Omega_{\rm m} = 3 \hat{m}^{-1} [(\Omega_{\rm n}^{-1})^\dagger - \Omega_{\rm n}^*] \hat{M} - (\Omega_{\rm n}^{-1})^T, 
\label{m}
\end{equation}
which must satisfy the requirement $\Omega_{\rm m}^{-1} = \Omega_{\rm m}^T$, i.e.,
\begin{equation}
3 \hat{M}^{-1}[ \Omega_{\rm n}^\dagger - (\Omega_n^*)^{-1}] \hat{m}-\Omega_{\rm n }^T = 3 \hat{M} [(\Omega_{\rm n}^{-1})^*-(\Omega_{\rm n}^*)^T] \hat{m}^{-1} - \Omega_{\rm n}^{-1}.
\label{-1=T}
\end{equation}

Assuming
\begin{equation}
A = \Omega_{\rm n}^T - \Omega_{\rm n}^{-1}, \qquad A \neq 0,
\end{equation}
Equation~(\ref{-1=T}) takes the form
\begin{equation}
3 \hat{M}^{-1} A^* \hat{m} + 3 \hat{M} A^* \hat{m}^{-1} - A = 0. \label{eqA}
\end{equation}

One can note that there is no matrix $A$ satisfying the relation (\ref{eqA}). 
Thus, the matrix $\Omega_{\rm nm}$ cannot be expressed by the form  (\ref{n+m}).

However, if one assumes that $\Omega_{\rm nm}$ is a diagonal matrix, then its form is defined explicitly as 
\begin{equation}
\Omega_{\rm nm} = {\rm diag} (\omega_1, \omega_2, \omega_3), \;\text{where }
\omega_i = \sqrt{1-\frac{1}{3} \frac{m_i}{M_i}}.
\end{equation}

\section{The effective masses of active neutrinos in standard and non-minimal~approximations}
\label{sec:mnu}

In the \textit{standard approach} using  the relations (\ref{M_M-lit}), (\ref{MN-lit}), (\ref{Theta-CI}) and $\Omega_{\rm m} \Omega_{\rm m}^T = I$, one can find
\begin{eqnarray}
m_\nu 
 \simeq  - M_D \theta^T  \simeq - \theta M_N \theta^T 
 \simeq  - \theta U_N^* \hat{M} U_N^\dagger \theta^T = - \Theta \hat{M} \Theta^T \simeq U_{\rm PMNS} \hat{m} U_{\rm PMNS}^T.
\label{eff_mass-R^-1}
\end{eqnarray}

In the \textit{non-minimal approximation}, 
the expression for $m_\nu$, (\ref{mnu}), with the help of (\ref{MN}), is
\begin{equation}
M_D \simeq  \left[ \theta - \frac{1}{3} (\theta^\dagger )^{-1} \right] M_N,
\end{equation}
and (\ref{eff-hat}) 
can be presented as
\begin{equation}
m_\nu \simeq - \left[ \theta - \frac{1}{3} (\theta^\dagger )^{-1} \right] U_N^* \hat{M} \Theta^T.
\end{equation}
 
Using (\ref{theta_nm}) and (\ref{Theta-RMD}), one can find 
\begin{eqnarray}
m_\nu  &\simeq & 
-i^2 U_\nu \sqrt{\hat{m}} \Omega_{\rm nm} \sqrt{\hat{M}^{-1}} U_N^T U_N^* \hat{M} \sqrt{\hat{M}^{-1}} \Omega_{\rm nm}^{-1} \sqrt{\hat{m}} U_\nu^T \nonumber \\
&\simeq & U_{\rm PMNS} \hat{m} U_{\rm PMNS}^T, 
\label{eff_mass-RMD}
\end{eqnarray}
where, as far as $|\theta| \ll 1$, one can assume $U_\nu \simeq U_{\rm PMNS}$, see (\ref{Upmns-nm}).


\begin{thebibliography}{999}


\bibitem{osc_1_1}
Pontecorvo, B.
Mesonium and anti-mesonium.
{\em Zh. Eksp. Teor. Fiz. (JETP)} {\bf 1957}, {\em 33}, 549--551.

\bibitem{osc_1_2}
Pontecorvo, B. 
Inverse beta processes and nonconservation of lepton charge.
{\em Sov. Phys. JETP} {\bf 1958}, {\em 7}, 172--173. 


\bibitem{osc_2_1}
Maki, Z.; Nakagawa, M.; Sakata, S.
Remarks on the unified model of elementary particles.
{\em Prog. Theor. Phys.} {\bf 1962}, {\em 28}, 870--880. 

\bibitem{osc_2_2}
Shrock R.E.
General theory of weak processes involving neutrinos. I. Leptonic pseudoscalar-meson decays, with associated tests for, and bounds on, neutrino masses and lepton mixing.
{\em Phys. Rev. D.} {\bf 1981}, {\em 24}, 1232--1274;
%

\bibitem{osc_2_3}
Shrock R.E.
General theory of weak processes involving neutrinos. II. Pure leptonic decays.
{\em Phys. Rev. D.} {\bf 1981}, {\em 24}, 1275--1309.

\bibitem{nuMSM_1}
Asaka, T.; Blanchet, S.; Shaposhnikov, M.
The $\nu$MSM, dark matter and neutrino masses.
{\em Phys. Lett. B} {\bf 2005}, {\em 631}, 151--156.

\bibitem{nuMSM_2}
Asaka, T.; Shaposhnikov, M.
The $\nu$MSM, dark matter and baryon asymmetry of the universe. 
{\em Phys. Lett. B} {\bf 2005}, {\em 620}, 17--26.

\bibitem{seesawI_1}
Minkowski, P. 
$\mu \to e \gamma$ at a rate of one out of $10^9$ muon decays?
{\em Phys. Lett. B} {\bf 1977}, {\em 67}, 421--428.

\bibitem{seesawI_2}
Gell-Mann, M.; Ramond P.; Slansky, R. 
Supergravity.
In {\em Proceedings of the Supergravity Stony Brook Workshop};
van Nieuwenhuizen, P., Freedman, D.Z., Eds.;   
North Holland Publ. Co.: Amsterdam, 1979 ; p. 315.

\bibitem{seesawI_3}
Yanagida, T. 
Horizontal gauge symmetry and masses of neutrinos.
In {\em Proceedinds of the Workshop on Unified Theories and Baryon Number in the Universe};
Sawada, O., Sugamoto A., Eds.; 
National Laboratory for High Energy Physics:
Tsukuba, Japan, 1979; pp. 95--99.

\bibitem{seesawI_4}
Mohapatra R.N.; Senjanovic, G. 
Neutrino mass and spontaneous parity nonconservation. 
{\em Phys. Rev. Lett.} {\bf 1980}, {\em 44}, 912--922.

\bibitem{dw}
Dodelson, S.; Widrow, L.M.
Sterile neutrinos as dark matter.
{\em Phys. Rev. Lett.} {\bf 1994}, {\em 72} 17--20.

\bibitem{7keV_th_1}
Dolgov, A.; Hansen, S.
Massive sterile neutrinos as warm dark matter.  
{\em Astropart.Phys.} {\bf 2002}, {\em 16}, 339--344. 

\bibitem{7keV_th_2}
Abazajian, K.; Fuller, G.M.; Tucker, W.H.
Direct detection of warm dark matter in the x-ray.  
{\em Astrophys. J.} {\bf 2001}, {\em 562}, 593--604.

\bibitem{7keV_exp_1}
Bulbul, E.; Markevitch, M.; Foster, A.; Smith, R.K.; Loewenstein, M.; Randall, S.W.
Detection of an unidentified emission line in the stacked x-ray spectrum of galaxy clusters.
{\em Astrophys. J.} {\bf 2014}, {\em 789}, 13--36.

\bibitem{7keV_exp_2}
Boyarsky, A.; Ruchayskiy, O.; Iakubovskyi, D.; Franse, J. 
Unidentified Line in x-ray spectra of the Andromeda
galaxy and perseus galaxy cluster.
{\em Phys. Rev. Lett.} {\bf 2014}, {\em 113}, 251301.

\bibitem{alehin}
Alekhin, S. et al. 
A facility to search for hidden particles at the CERN SPS: the SHiP physics case.
{\em Rept. Prog. Phys.} {\bf 2016}, {\em 79}, 124201.

\bibitem{1-50keV}
Boyarsky, A.; Ruchayskiy, O.; Shaposhnikov, M. 
The role of sterile neutrinos in cosmology and astrophysics.
{\em Ann. Rev. Nucl. Part. Sci.} {\bf 2009}, {\em 59}, 191--214. 

\bibitem{2007_Shap_sym}
Shaposhnikov, M.
A possible symmetry of the $\nu$MSM. 
{\em Nucl. Phys. B} {\bf 2007}, {\em 763}, 49--59.

\bibitem{pmns_3}
Pontecorvo, B.  
Neutrino experiments and the question of leptonic-charge conservation.  
{\em Sov.Phys. JETP} {\bf 1968}, {\em 26}, 984--988. 

\bibitem{merle}
Merle A.; Rodejohann, W.  
The elements of the neutrino mass matrix: Allowed ranges and
implications of texture zeros.
{\em Phys. Rev. D} {\bf 2006}, {\em 73}, 073012. 

\bibitem{0nubb}
CUORICINO collaboration. 
Results from a search for the $0 \nu \beta \beta$-decay
of 130 Te.
{\em Phys. Rev. C)} {\bf 2008}, {\em 78}, 035502. 

\bibitem{GA}
Fogli, G.L. 
et. al. 
Global analysis of neutrino masses, mixings and phases: entering the era of leptonic CP violation searches.
{\em Phys. Rev. D} {\bf 2012}, {\em 86}, 013012.  

\bibitem{Plank}
Planck collaboration. 
Planck 2018 results. VI. Cosmological parameters.
{\em Astron. Astrophys.} {\bf 2020}, {\em 641}, A6; 
Erratum in: 
{\em Astron. Astrophys.} {\bf 2021}, {\em 652}, C4. 

\bibitem{drewes_22}
Drewes, M.; Georis, Y.; Hagedorn, C.; Klaric, J. 
Low-scale leptogenesis with flavour and CP symmetries.
{\em JHEP} {\bf 2022}, {\em 12}, 044. 

\bibitem{nuMSM_lightest}
Boyarsky, A.; Neronov, A.; Ruchayskiy, O.; Shaposhnikov, M.
The masses of active neutrinos in the $\nu$MSM from x-ray astronomy. 
{\em JETP Lett.} {\bf 2006}, {\em 83}, 133--135.

\bibitem{MN<1eV_1}
de Gouvea, A.
See-saw energy scale and the LSND anomaly.
{\em Phys. Rev. D} {\bf 2005}, {\em 72}, 033005.

\bibitem{MN<1eV_2}
de Gouvea, A.; Jenkins, J.; Vasudevan, N.  
Neutrino phenomenology of very low-energy seesaws. 
{\em Phys. Rev. D} {\bf 2007}, {\em 75}, 013003.

\bibitem{HNLser_dir1}
Arguelles, C.A.; Foppiani, N.; Hostert, M.  
Heavy neutral leptons below the kaon mass at hodoscopic detectors.
{\em Phys. Rev. D} {\bf 2022}, {\em 105}, 095006.

\bibitem{HNLser_dir2}
Kelly K.J.; Machado, P.A.N.  
MicroBooNE experiment, NuMI absorber, and heavy neutral leptons.
{\em Phys. Rev. D} {\bf 2021}, {\em 104}, 055015.

\bibitem{HNLser_cosm1}
Hernandez, P.; Kekic, M.; Lopez-Pavon, J.
$N_{\rm eff}$ in low-scale seesaw models versus the
lightest neutrino mass. 
{\em Phys. Rev. D} {\bf 2014}, {\em 90}, 065033.

\bibitem{HNLser_cosm2}
Mastrototaro, L.; Serpico, P.D.; Mirizzi, A.; Saviano, N. 
Massive sterile neutrinos in the early Universe: From thermal decoupling to cosmological constraints.
{\em Phys. Rev. D} {\bf 2021}, {\em 104}, 016026.

\bibitem{HNLser_cosm3}
Boyarsky, A.; Ovchinnikov, M.; Ruchayskiy, O.; Syvolap, V. 
Improved big bang nucleosynthesis constraints on heavy neutral leptons.
{\em Phys. Rev. D} {\bf 2021}, {\em 104}, 023517.

\bibitem{lep_univ}
Asaka, T.; Eijima, S.; Takeda, K.  
Lepton universality in the $\nu$MSM.
{\em Phys. Lett. B} {\bf 2015}, {\em 742}, 303--309.

\bibitem{MN>100}
Asaka, T.; Eijima, S. 
Direct search for right-handed neutrinos and neutrinoless double beta decay.  
{\em PTEP} {\bf 2013}, {\em 11}, 113B02.

\bibitem{canetti_13}
Canetti, L.; Drewes, M.; Frossard, T.; Shaposhnikov, M.
Dark matter, baryogenesis and neutrino oscillations from right-handed neutrinos.
{\em Phys. Rev. D} {\bf 2013}, {\em 87}, 093006.

\bibitem{mp}
Mohapatra, R.N.; Pati J.C.
A natural left-right symmetry.
{\em Phys. Rev. D} {\bf 1975}, 2558


\bibitem{casas_ibarra}
Casas, J.; Ibarra, A.  
Oscillating neutrinos and $\mu \to e, \gamma$,
{\em Nucl. Phys. B} {\bf 2001}, {\em 618}, 171.


\bibitem{white}
Adhikari, R.; Agostini, M.; Ky, N.A.; Araki, T.; Archidiacono, M.; Bahr, M.; Baur, J.; Behrens, J.; Bezrukov, F.; Dev, P.B.; et al.
A White Paper on keV Sterile Neutrino Dark Matter. 
{\em JCAP} {\bf 2017}, {\em 01}, 025.



\bibitem{bilenky}
Bilenky, S.M.; Petcov, S.T.
Massive neutrinos and neutrino oscillations. 
{\em Rev. Mod. Phys.} {\bf 1987}, {\em 59}, 671;
Erratum in: {\bf 1988}, {\em 60}, 575;
Erratum in: {\bf 1989}, {\em 61}, 169.



\bibitem{molinaro}
Ibarra, A.; Molinaro, E.; Petcov, S.
TeV scale see-saw mechanisms of neutrino mass generation, the Majorana nature of the heavy singlet neutrinos and $(\beta \beta) 0 \nu$-decay.
{\em J. High Energ. Phys.} {\bf 2010}, {\em 09}, 108.

\bibitem{DK_jetp}
Dubinin, M.N.; Kazarkin D.M. 
Improved cosmological bounds for a fine-tuned see-saw mechanism of keV sterile neutrinos. \emph{arXiv} \textbf{2022}, arXiv:2206.05186.

\bibitem{das_22}
Das, P.; Das, M.K.; Khan, N. 
Five-zero texture in neutrino-dark matter model within the framework of minimal extended seesaw.
{\em Nucl. Phys. B} {\bf 2022}, {\em 980}, 115810.


\bibitem{DFK}
Dubinin, M.N.; Fedotova, E.Y.; Kazarkin, D.M.  
Cosmological constraints for three generations see-saw mechanism of keV sterile neutrinos.
{\em Phys. Part. Nucl. Lett.} {\bf 2023}, {\em 20}
(in press).

\bibitem{grounau}
Gronau, M.; Leung, C. N.; Rosner, J. L.
Extending limits on neutral heavy leptons.
{\em Phys. Rev. D} {\bf 1984}, {\em 29}, 2539

\bibitem{DK_lep_un}
Dubinin, M.N.; Kazarkin, D.M. 
Lepton universality in a model with three generations of sterile Majorana neutrinos. arXiv \textbf{2022}, arXiv:2212.11310.

\bibitem{drewes_LNV}
Drewes, M.; Klarić, J.; Klose, P.
On lepton number violation in heavy neutrino decays at colliders.
{\em JHEP} {\bf 2019}, {\em 11}, 032.

\bibitem{shap_leptogenesis}
Shaposhnikov, M. 
The $\nu$MSM, leptonic asymmetries, and properties of singlet fermions.
{\em JHEP} {\bf 2008}, {\em 8}, 008.  



\end{thebibliography}
\end{document}